\documentclass[12pt]{article}
\pdfoutput=1
\usepackage{epsfig}
\usepackage{amsfonts}
\usepackage{amssymb}
\usepackage{xcolor}
\usepackage{amsmath}
\usepackage{upgreek}
\usepackage{graphicx}
\usepackage{cite}
\usepackage{calligra}
\usepackage{caption}
\usepackage{subcaption}

\usepackage{cmap}					% поиск в PDF
\usepackage{mathtext} 				% русские буквы в формулах
\usepackage[english]{babel}	% локализация и переносы
\usepackage{indentfirst}
\usepackage{hyperref}

\begin{document}
\newcommand{\nc}{\newcommand}
\nc{\beq}{\begin{equation}} \nc{\eeq}{\end{equation}}
\nc{\beqa}{\begin{eqnarray}} \nc{\eeqa}{\end{eqnarray}}
\nc{\R}{{\cal R}}
\nc{\A}{{\cal A}}
\nc{\K}{{\cal K}}
\nc{\B}{{\cal B}}
\begin{center}

{\bf \Large Leading all-loop quantum contribution to the effective potential in the inflationary  \\[0.3cm] cosmology} \vspace{1.0cm}

{\bf \large   D.I. Kazakov$^{1,2}$, R. M. Iakhibbaev$^{1}$ and D.M. Tolkachev$^{1,3}$} \vspace{0.5cm}

{\it $^1$Bogoliubov Laboratory of Theoretical Physics, Joint
Institute for Nuclear Research, Dubna, Russia, 141980\\
$^2$Moscow Institute of Physics and Technology, Dolgoprudny, Russia, 141701\\[0.1cm]
$^3$Stepanov Institute of Physics, Minsk, Belarus, 220072}

\vspace{0.5cm}

\abstract{In this paper, we have constructed quantum effective potentials and used them to study slow-roll inflationary cosmology. We derived the generalised RG equation for the effective potential in the leading logarithmic approximation and applied it to evaluate the potentials of the $T^2$ and $T^4$-models, which are  often used in modern models of slow-roll inflation.  We found  that while the one-loop correction strongly affects the potential, breaking its original symmetry, the contribution of higher loops smoothes the behaviour of the potential. However, unlike the $\phi^4$-case, we found that the effective potentials preserve spontaneous symmetry breaking when summing all the leading corrections.
We calculated the spectral indices $n_s$ and $r$ for the effective potentials of both models and found that they are consistent with the observational data for a wide range of parameters of the models.}

\end{center}

\section*{Introduction}

The inflationary model based on the assumption of the existence of an accelerated expansion stage in the early Universe, became one of the foundations of modern cosmology \cite{Guth1980, Starobinsky1980}. This area  is rapidly developing  because it explains the observed flatness, homogeneity and isotropy of the observable Universe. The advantage of the inflationary theory is also related to its successful explanation of the peculiarities of the CMB spectrum \cite{Mukhanovbook}.

The most common way to realise the accelerated expansion is to use a scalar field (inflaton or inflationary field) with self-interaction and to consider the solution of the field equations in slow-roll regime. Obtaining an inflaton potential is a rather difficult problem, since at present the observational data give a wide scope for consideration of a large number of potentials  \cite{Martin}. Recently, $\alpha$-attractor type models in inflationary cosmology have aroused considerable interest, mainly because they are justified in supergravity and satisfy observational data.  \cite{Kallosh, Galante, Kallosh2013}. Therefore, their (or their possible modifications) detailed study in the context of inflation is quite an interesting task \cite{Galante}.

In quantum theory, the classical inflaton potential acquires quantum loop corrections. Such corrections, first considered in the paper by Coleman and Weinberg ~\cite{CW}, can change the form of the potential and lead to the appearance of a new vacuum, which means spontaneous symmetry breaking takes place. The inflationary model based on the  $\phi^4$ theory with a  one-loop quantum correction was considered in ~\cite{Rehman:2008qs}. Within the framework of this model, the old theory of inflation with tunneling  \cite{Guth1980} and then the new inflation in models with slow rolling \cite{Rehman:2008qs} was formulated. Note, however, that taking into account only one-loop corrections is not always justified. As was shown in the original work ~\cite{CW}, summing the leading terms with the renormalization group significantly changes the behavior of the potential. 
Hence, the construction of an effective potential that takes into account all leading corrections is a current task.

However, the all-loop effective potentials are not often used in cosmology. Existing literature (see e.g.\cite{Elizalde:1993ee, Elizalde:1993qh}) deals with  effective potentials in renormalisable cases when the formalism of the usual renormalisation group formalism is applicable. At the same time, the most popular  cosmological potentials from the point of view of quantum field theory, are non-renormalizable. This leads to all the problems typical of non - renormalizable theories such as infinite arbitrariness in normalisation of counter terms, absence of standard renormalization group (RG) equations for summation of leading corrections, scheme dependence of results of calculations, etc. However, if one considers only the leading quantum corrections (the leading logarithmic approximation), these problems can be avoided. Indeed, the leading divergences (they are also the leading orders of  $\log\phi^2$) are universal, they do not depend on the arbitrariness of the subtraction procedure (they are not scheme-dependent). Moreover, this property is independent of the renormalizability or non-renormalizability of the theory. Thus, one can try to calculate the main quantum corrections to the effective potential and apply the obtained expression for the description of inflation, ignoring the  main problems of non-renormalizable theories so far.  

In the present paper we take into account the leading quantum corrections to the classical potential. Considering an arbitrary scalar potential, we derive a generalised RG equation which sums all leading logarithmic corrections to the potential. This way, we derive an effective potential and based on it we analyse the cosmological parameters that are obtained from this modified potential.  In the following, we consider the potential for the so-called $T$-model \cite{Kallosh2013}, which can be represented as 
\begin{equation}V=\tanh^{2n} (\phi/(\sqrt{6\alpha M_{Pl}^2})),
\end{equation}
where $\phi$ is the inflaton field, $M_{Pl}$ is Planck's mass, but $\alpha$ is a free parameter, $n=1,2$. A qualitative analysis of the behaviour of quantum corrections to such potentials was carried out in \cite{Kallosh:2016gqp}, and in this paper we will be able to check their conclusions about flat asymptotic behaviour of the one-loop corrected potential. 

Based on the obtained numerical solutions of the RG equation, effective potentials and cosmological parameters, we can understand at under parameters the effective potential satisfies the observational data. 

\section{Single-field model of slow-roll inflationary scenario: setup and observables}

The starting point in inflationary cosmology with a single field is the scalar-tensor theory, with the action that can be presented as \cite{Martin, Motohashi}:
\begin{equation}
    S = \int d^4x\sqrt{-g} \left[ \frac{M_{pl}^2}{2}R + \frac{1}{2}g^{\mu\nu}\partial_\mu \phi \partial_\nu \phi - V(\phi)  \right],
\end{equation}
where $R$ is a scalar curvature, $M_{Pl}=8 \pi G$ is Planck's mass, $\phi$ is the inflation scalar field, and $V(\phi)$ is self-interaction term. In the Friedman-Lemetre-Robertson-Walker metric, the equations of motion for the above model can be written in the form \cite{Weinberg:2008zzc}
\begin{equation}
    3M_{pl}^2H^2 = \frac{\dot{\phi}^2}{2} + V,\label{feq1}
\end{equation}
\begin{equation}
    -2M_{pl}^2\dot{H} = \phi^2,\label{feq2}
\end{equation}
\begin{equation}
    \ddot{\phi} + 3H \dot{\phi} + \frac{\partial V}{\partial \phi} = 0. \label{feq3}
\end{equation}
Here $H=\dot{a}/a$ is Hubble's constant and $a(t)$ is a scale factor.

To describe the evolution of the background, it is useful to introduce cosmological parameters of the Hubble flow:
\begin{equation}
    \epsilon_n=\frac{d \log \epsilon_{n-1}}{d N_e},
\end{equation}
here $\epsilon_0=H_{0}/H$, with $H_0$ being initial Hubble's constant, $N_e$ is the ratio of the scale factor at the end of inflation to the factor at its beginning: $N_e=\ln(a/a_0)$ (also $N_e$ is called the number of $e$-foldings). Cosmological parameters are responsible for the nature of inflation and reflect the shape of the inflation potential. Inflation is an accelerated expansion, and the condition for such an expansion can be written as $\epsilon_1<1$, and the end of inflation corresponds to $\epsilon_1=1$. Under the condition of correctness of the slow roll approximation, we can consider  that $\epsilon_n \ll 1$, so that the following estimation of the cosmological parameter becomes valid:
\begin{equation}
    \epsilon_1  \simeq  \frac{1}{2} \left(\frac{V'}{V}\right)^2 = \frac{1}{2 M_{pl}}\left(\frac{d \phi}{d N_e}\right)^2.
    \label{epseq}
\end{equation}

The condition $\epsilon=1$ with the help of (\ref{epseq}) can be rewritten in terms of the Klein-Gordon-Fock equation \cite{Martin}:
\begin{equation}
    \frac{1}{M_{pl}} \left(\frac{d \phi}{d N_e}\right)^2=\frac{d \ln V}{d N_e}. \label{eqfN}
\end{equation}
This equation can be integrated to give an explicit expression for the number of $e$-folds \cite{Martin}:
\begin{equation}
   N_e-N_{0}= -\frac{1}{M_{pl}^{2}}\int_{\phi_{0}}^{\phi}\frac{V(x)}{V_{x}(x)}dx.
\end{equation}
Based on observational evidence, we can also say that the lower limit for expansion of the Universe should be a number of 50 to 60 $e$-folds. 
The observable characteristics of inflation are the spectral indices: CMB tilt of scalar perturbations \cite{Kallosh2013, Martin}
\begin{equation}
    n_s=1-2\epsilon_1+\epsilon_2, 
\end{equation}
the tilt of tensor perturbation \cite{Martin}: 
\begin{equation}
    n_t=-2 \epsilon_1.
\end{equation}
The CMB tensor-to-scalar ratio is \cite{Martin}:
\begin{equation}
    r=16 \epsilon_1.
\end{equation}
According to the relic radiation data obtained from the  BICEP/PLANCK \cite{BICEP,PLANCK}, the observed values for these quantities are as follows:
\begin{equation}
    r<0.036, ~ n_s = 0.9649 \pm 0.0042.
\end{equation} 
The $n_s$-$r$ plot generally allows one to discriminate models and compare them with observational data.

In this paper we consider the T-model potential, which we will rewrite as follows \cite{Ketov, Galante}
\begin{equation}
    V(\phi)=g \tanh^{2 n}\left(\phi \omega/\sqrt{6 \alpha}\right),  \label{potT}
\end{equation}
where $\omega =  M_{Pl}^{-1}$, $g$ is an appropriate inflation scale, and $\alpha$ we put equal to unity. In the following, we will restrict ourselves to two cases of T-model: $n=1$ (for convenience, we will call it the $T^2$-model) and $n=2$ ($T^4$-model). According to the above \eqref{eqfN}, above for cosmological parameters, one can easily find the exact analytical formula for the dependence of  $\phi$ on $N_e$ for T-models \cite{Martin}:
\begin{equation}
    \phi^{(n)} (N_e)={\frac{\sqrt{6}}{2 \omega} }\operatorname{arccosh}\left(\sqrt{1+4/3 n^2}+4/3 ~n \Delta N\right),
\end{equation}
where $\Delta N=N_f-N_e$, and $N_f$ can be found from the inflation ending condition $\epsilon_1=1$. Due to the simple form of the  $\phi(N)$ dependence the corresponding spectral parameters of the Hubble flow can be obtained. Thus, the T-model is quite convenient, since all parameters in it are represented in analytical form. In a general form, the functions and spectral parameters for T-models in arbitrary form are presented in ref. \cite{Martin}.

\section{Effective potential in scalar field theory with\\ arbitrary potential}	

The effective potential is defined as part of the effective action without derivatives.
The direct way to find the effective potential $V_{eff}(\phi)$ by perturbation theory is to calculate the sum of one-particle irreducible vacuum diagrams obtained using Feynman's rules derived from  the shifted action $S[\phi+\widehat \phi]$,  where $\phi$ is the classical field obeying the equation of motion and $\widehat \phi(x)$ is the quantum field  over which integration is performed ~\cite{EA}. This means that one has to consider the 1PI vacuum diagrams with propagators containing an infinite number of insertions $v_2(\phi)\equiv \frac{d^2V(\phi)}{d\phi^2}$, which act like a mass depending on the field $\phi$: $m^2(\phi)=gv_2(\phi)$. The vertices are also obtained from the expansion of the potential $V(\phi+\widehat \phi)$ by the quantum field $\widehat\phi$. After that, the effective potential is constructed as a perturbation expansion by the coupling constant $g$
\beq
V_{eff}=g\sum_{n=0}^\infty (-g)^n V_n,
\eeq
where $V_0=V$ is the initial classical potential.

We choose a dimensional regularisation to control UV divergence in loop integrals, taking the dimension of spacetime $D=4-2 \epsilon$.
Proceeding in this way, in the one-loop case, we obtain the quantum correction~\cite{we2023}:
\begin{equation}
V_1=  \frac{1}{16\pi^2}\frac 14\frac{v_2^2}{\epsilon} \left(\frac{\mu^2}{m^2}\right)^\epsilon \to \frac{1}{16\pi^2}\frac{v_2^2}{4}\left(\frac 1\epsilon +\log \frac{\mu^2}{m^2}\right), \ \ \  m^2=gv_2(\phi).
\end{equation}
 Further, the singular part $\sim 1/\epsilon$ is removed by introducing of the UV counterterms, and the finite part $\sim \log(gv_2(\phi))$ gives the contribution to the effective potential. Note that the coefficient of the pole $1/\epsilon$ exactly coincides with the coefficient of the logarithm.  This property is also preserved in higher loops: the coefficient of the leading pole $ 1/\epsilon^n$ coincides with the coefficient of the leading  logarithm $\log^n(gv_2(\phi))$.
Hence, the task is to find the coefficients of the leading poles, which, in turn, due to the special features of the ${\cal R}$-operation is determined by one-loop diagrams~\cite{we2023}.

The next step is to obtain recurrence relations connecting the leading divergences in subsequent loops. They obviously follow from the local structure of $\mathcal{R}$-operations~\cite{we2023}.
Denoting the singular part of the effective potential (coefficient at the leading pole $1/\epsilon^n$) in the n-th order of perturbation theory by $\Delta V_n$, one can obtain the following recurrence relation~\cite{we2023}
\beq
n \Delta V_n=  \frac 14 \sum_{k=0}^{n-1} D_2  \Delta V_k D_2  \Delta V_{n-1-k}, \ \ \, n\geq1, \ \   \Delta V_0=V_0 ,  \label{rec2}
\eeq
where $D_2$ is the second derivative by the field $\phi$. 
This recurrence relation allows one to compute all leading divergences $\Delta V_n$ for an arbitrary potential $V$ in an algebraic way without computing diagrams.

To sum up the leading divergences (or, what is the same, the leading logarithms in the finite part), it is convenient to pass to the differential equation for the sum of series  
\beq
\Sigma(z,\phi)=\sum_{n=0}^{\infty} (-z)^n\Delta V_n(\phi),\label{sum}
\eeq
where $z=g/\epsilon$.
 Indeed, multiplying Eq.(\ref{rec2}) by the factor $(-z)^n$ and summing over n from $n=2$ to $\infty$, we obtain the differential equation for the function $\Sigma(z,\phi)$, 
 \beq
 \frac{d \Sigma}{dz}=-\frac 14 (D_2 \Sigma)^2, \ \ \Sigma(0,\phi)=V_0(\phi). \label{RG}
\eeq
This is the desired generalised RG equation for the effective potential in the approximation of leading logarithms.  For renormalizable, interaction it reduces to the usual RG equation \cite{we2023}. To receive an effective potential in the solution of this equation, one should replace the pole by $\epsilon$ with the corresponding logarithm: 
\beq
V_{eff}(g,\phi)=\Sigma(z,\phi)|_{z\to -\frac{g}{16\pi^2}\log{gv_2/\mu^2}}.\label{effpot}
\eeq
Recall that here  $v_2(\phi)\equiv \frac{d^2V_0(\phi)}{d\phi^2}$.

Note that Eq.(\ref{RG}) is a partial derivative equation, and the function $\Sigma(z,\phi)$ depends on two variables: $z$ and $\phi$. In general, it is usually more convenient to reduce this function to dimensionless variables. In some special cases, (\ref{RG}) can be reduced to an ordinary differential equation, but still one has to use numerical methods to solve it \cite{we2023}.

\section{Inflation and effective potentials}
\subsection{$T^2$-model}

Consider the theory with the potential (\ref{potT}) for $n=1$, which corresponds to the $T^2$-model:
\begin{equation}
    V= g \tanh^2 \left(\frac{\phi \omega}{\sqrt{6}}\right).
\end{equation}
Further, to simplify equation (\ref{RG}) it is convenient to represent the function $\Sigma$ in dimensionless variables  $x=z\omega^4$ and  $y=\tanh^2(\phi\omega/\sqrt{6})$, because the loop decomposition of this function can be represented as polynomials on tangents. Then
$$\Sigma(z \omega^4, \tanh^2(\phi \omega/\sqrt{6}))\equiv S(x,y).$$
The obtained function is in some way an analogue of an arbitrary  function $F(\tanh(\frac{\phi}{\sqrt{6}}))$ in the theory of conformal chaotic inflation \cite{Kallosh2013}. But in our case this function is restricted by the RG equation \eqref{RG} and it contains an additional regularisation parameter $\mu^2$. 

Renaming the variables and functions in the original equation \eqref{RG},
the generalised RG-equation for such a potential can be written as:
\begin{equation}
    \frac{\partial}{\partial x}S(x, y)=-\frac{1}{4} \left( \omega^2 \frac{\partial^2}{\partial \phi^2}S(x, y(\phi)) \right)^2.
\end{equation}
Such change of variables considerably simplifies the form of the equation, so that for the $T^2$-model the generalised RG-equation, after the transformation from the derivative over $\phi$ to the derivative over $y$, has the following form:
\begin{equation}
    \frac{1}{36} (y-1)^2 \left((1-3 y) S_y-2 (y-1) y S_{yy}\right)^2=-S_x. \label{eq}
\end{equation}
For brevity, the lower indices are introduced here to denote the corresponding derivative of the function $S$. The initial and boundary conditions for the equation can be defined as:
\begin{equation}
S(0,y)=y, ~ S(x,1)=1, ~ S_x(x,1)=0,
\end{equation}
because the potential in the limit $g=0$ must coincide with the classical potential, as stated in (\ref{RG}) and as $\phi \rightarrow \pm \infty$ must reach a constant. The exact solution of Eq. (\ref{eq}) cannot be represented analytically: it is a complex two-dimensional surface, and to find an effective potential, we have to perform a substitution $z\to -\frac{g}{16\pi^2}\log{g v_2/\mu^2}$, and in the case of the $T^2$-model   \cite{we2023}
\begin{equation}
    v_2=-\frac{1}{3} \omega ^2 \left(\cosh \left(\sqrt{\frac{2}{3}} \omega  \phi \right)-2\right) \text{sech}^4\left(\frac{\omega  \phi }{\sqrt{6}}\right).
\end{equation}
That is, this substitution is equivalent to the choice of a particular curve on a two-dimensional surface given by the solution of Eq. (\ref{eq}). 
The numerical solution of the equation can be carried out using Euler's backward differentiation method \cite{stiff} or built-in methods in \texttt{Wolfram} \texttt{Mathematica}. 

The initial potential has flat asymptotics, so it is often used in models of chaotic inflation, as it was indicated earlier due to the attractor properties of solutions (solutions of Friedmann equations do not depend on initial conditions and shrink into a stable limit cycle in the phase space). The solutions of the generalised RG-equation have the same asymptotics but can modify the behaviour of the potential in inflationary cosmological models leading to different slow roll scenarios and creating a different dynamical picture of solutions. More specifically, one can study the form of the total quantum effective potential at different parameters $g, \mu^2$. And since they are related, it is possible to restrict them by modifying one of them, say, fixing \footnote{The scale of inflation is considered to be equal to about $10^{-5}M_{Pl}$} $g~\sim 1$ , and changing $\mu$ to satisfy the conditions of convergence. The convergence conditions of the original series for the potential (\ref{sum}) are represented by the following expressions \cite{we2023}:
\begin{equation}
   \log g\frac{\omega^2 (2-\cosh(\phi \omega))}{\mu^2 \cosh^4(\phi \omega)}>1, ~ \frac{g \omega^4}{16\pi^2}  < 1,
\end{equation}
they set the bounds of applicability of the approximation of the leading logs for the effective potential.

\begin{figure}[h!]
\begin{center}
\includegraphics[scale=0.6]{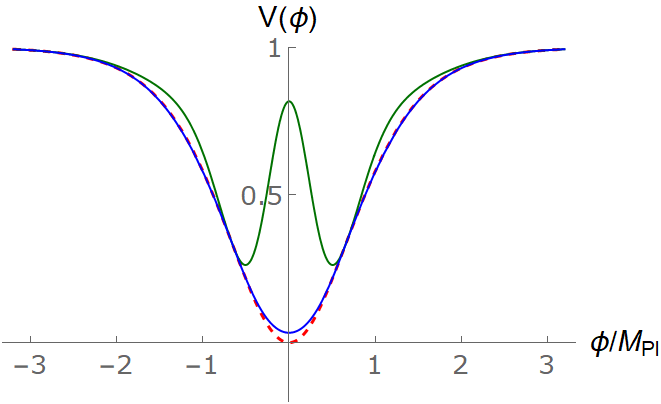}
\includegraphics[scale=0.6]{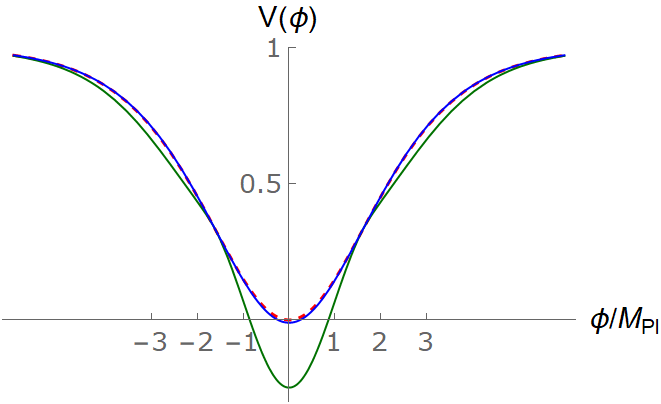}
\caption{Comparison of the classical potential (red dashed line), the one-loop correction (green solid line), and the RG effective potential (blue solid line) at $g \sim 1, ~ \mu^2 \sim 10^{-2}$ (upper plot) and $g \sim 1, ~ \mu^2 \sim 10$ (lower plot)} \label{fig:oneloop}
\end{center}
\end{figure}

%\begin{figure}[h!]
%\begin{center}
%\includegraphics[scale=0.7]{mu2ll1.png}
%\caption{Comparison of classical potential (red dashed %line), classical potential with one-loop contribution (green solid line) and resummed effective potential (blue solid line) at $g \sim 1, ~ \mu^2 \sim 10$} \label{oneloop1}
%\end{center}
%\end{figure}

We compare the behaviour of the resummed potential with a one-loop contribution which has the following form
\begin{equation}
    V_{1loop}(\phi)= V(\phi)+\frac{g^2}{16 \pi^2}\frac{1}{4}v_2^2 \log \left(\frac{g v_2}{\mu^2}\right).
\end{equation}

In fact, the $T^2$-model ($T^4$-model either) is called an attractor because the phase portrait of the Friedman equations (\ref{feq1}-\ref{feq3}) represents a limit cycle (or attractor), i.e., the general dynamics of this system does not depend on the initial conditions. So introducing additional corrections to this potential can affect the dynamical properties, and there can appear the Hopf bifurcation picture \cite{Toappear}. 

From Fig.\ref{fig:oneloop} one can see that the one-loop potential strongly modifies the classical part of $V(\phi)$ causing a maximum (or minimum at $\mu^2 >1$) at $\phi=0$, while the all-loop quantum effective potential smoothes the perturbation from the first loop contribution. One can conclude from this observation that the quantum effective potential is more stable in contrast to the one-loop-corrected potential. We also note that for small $\mu$, a situation may occur where the contributions of all loops are unable to flatten the maximum near $\phi=0$.

\begin{figure}[h!]
\begin{center}
\includegraphics[scale=0.7]{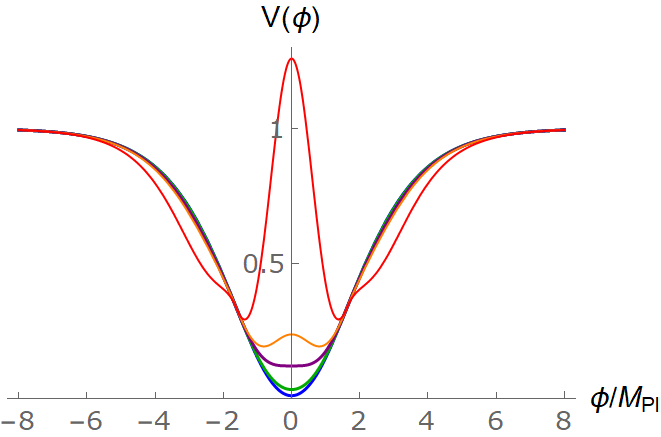}
\caption{Comparison of the behaviour of different potentials with all loop corrections at $g \sim 1$.  Blue line --- $\mu^2 \sim 0.6$, green $\mu^2 \sim 10^{-2}$, purple $\mu^2 \sim 10^{-10}$, orange $\mu^2 \sim 10^{-16}$, red $\mu^2 \sim 10^{-36}$} \label{diffg}
\end{center}
\end{figure}

The relevant potentials at different $\mu$ are shown in Figure \ref{diffg}. Spontaneous symmetry breaking with a maximum near zero is observed and, starting from certain extremely small $\mu^2 \sim 10^{-80}$, additional minima can arise which correspond to a false vacuum.

The varying behaviour of the potential can find its application in slow rolling. It is convenient to depict the whole set of cosmological parameters to demonstrate how inflation changes for different values of $\mu^2$. Using the formulas from the first chapter and numerical calculations, we plot the value of $\epsilon_1$  as a function of $\phi$. Based on these plots, one can conclude that the occurrence of extra minima for some values of  $\mu^{2}$  does not strongly affect the slow roll  behaviour, although, as one can see,  the smaller $\mu^2$, the shorter is the inflation (e.g., for the classical $T^2$-model the inflation ends at $\phi_{end} \omega \simeq 1.208$, while for $\mu^2 \sim 10^{-3}$ the inflation ends at $\phi_{end} \omega \simeq 1.061$).  Thus, the slow rolling is realised within the observational data for rather wide bounds on $\mu^2$.

\begin{figure}
\centering
\begin{subfigure}{.5\textwidth}
  \centering
  \includegraphics[width=.92\linewidth]{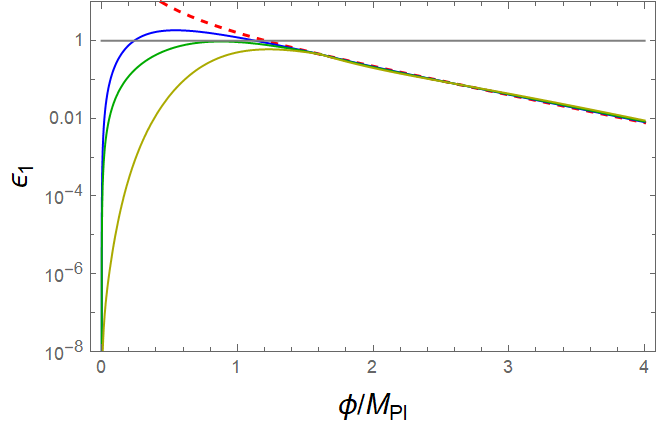}
  \caption{}
  \label{fig:sub1}
\end{subfigure}%
\begin{subfigure}{.5\textwidth}
  \centering
  \includegraphics[width=.92\linewidth]{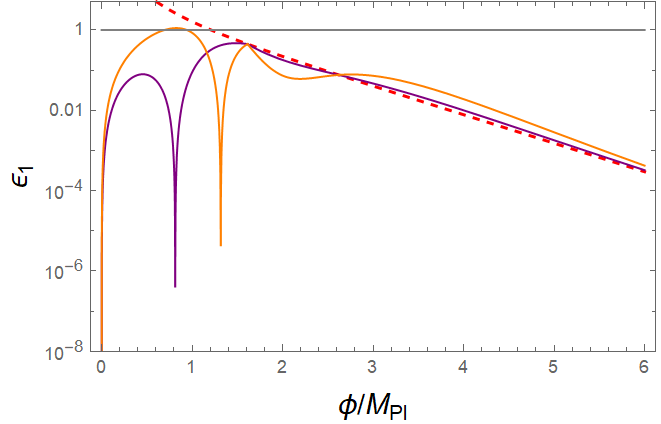}
  \caption{}
  \label{fig:sub2}
\end{subfigure}
\caption{Plots for the Hubble flow parameter $\epsilon_1$   In Fig. \ref{fig:sub1} the blue line  --- $\mu^2 \sim 10^{-2}$, green line  --- $\mu^2 \sim 10^{-8}$, yellow line  --- $\mu^2 \sim 10^{-10}$ ,  at the \ref{fig:sub2} purple line $\mu^2 \sim 10^{-16}$, orange one $\mu^2 \sim 10^{-56}$. The red dashed line corresponds to the classical potential and the grey line indicates the end of inflation}
\label{epsdiff}
\end{figure}

Interestingly, for the all-loop effective potential with $\mu^2 \leq 10^{-5}$, the situation arises when slow-roll inflation can become eternal (for the one-loop potential perpetual inflation occurs already with $\mu^2 \leq 10^{-1}$) \cite{Greenwood:2021uuj,Guth2007ng}. This is due to spontaneous symmetry breaking and the fact that the maximum of the effective potential is less than unity in our notation. That is, the behaviour of the potential deteriorates when $\mu^2$ decreases, since false vacua emerge. The existence of false vacua leads to the problem of phase transitions in the early Universe \cite{GuthWeinberg}, although on the other hand, these features can change the power spectrum of gravitational waves at certain scales, which may be useful for explaining primordial black holes production \cite{Dalianis}. The question of how exactly tunnelling through potential barriers can occur was studied in the literature \cite{GuthWeinberg, WeinbergInstab}, and is beyond the scope of our consideration (as well as the analysis of the postinflationary reheating phase in these models). A qualitative study of the reheating phase for corrected $T$-model potential was made in \cite{Kallosh:2016gqp}.

 \begin{figure}[h]
\begin{center} 
\includegraphics[scale=0.7]{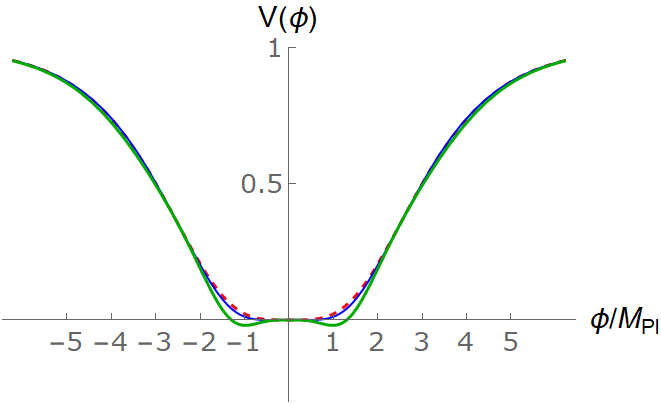}
\caption{Classical potential $\tanh^4$ (red dashed line), single-loop (green solid line) and full effective (blue solid line) potentials at $\mu^2 \sim 10^{-80} $, the purple solid line corresponds to the quantum effective potential at $\mu^2 \sim 10^{-80} $ }\label{t4}
\end{center}
\end{figure}

\subsection{$T^4$-model}

Similar to the case of the $T^2$-model, we can consider $T^4$-model with the potential 
\begin{equation}
    V = g \tanh^4( \frac{ \phi \omega }{\sqrt{6}} ).
\end{equation}
For this potential, one can also write down a generalised RG equation with the same initial and boundary conditions and then construct the effective potential.  The initial differential equation (\ref{RG}) can be written in the form
\begin{equation}
    \frac{\partial}{\partial z  }\Sigma(z \omega^4, \tanh^4(\phi \omega/\sqrt{6}))=-\frac{1}{4} \left(\frac{\partial^2}{\partial \phi^2}\Sigma(z \omega^4, \tanh^4(\phi \omega/\sqrt{6}))\right)^2
\end{equation}
After changing the variables to dimensionless $x=z \omega^4$, $y=\tanh^4(\phi \omega / \sqrt{6})$ and the desired function as $\Sigma(z \omega^4, \tanh^4(\phi \omega/\sqrt{6}))=S(x,y)$ the differential generalised RG-equation, after passing from the derivative of $\phi$ to the derivative on $y$, will have the following form:
\begin{equation}
    \frac{1}{9} \left(y^{1/2}-1\right)^2 y \left(\left(5 y^{1/2}-3\right)  S_{y}+4 y \left(y^{1/2}-1\right) S_{yy}\right)^2=-S_x \label{eqt4}, 
\end{equation}
with the boundary and initial conditions in the same form as for the $T^2$-model:
\begin{equation}
    S(0,y)=y, ~ S(x,1)=1,~ S_x(x,1)=0.
\end{equation}
Equation (\ref{eqt4}) looks more complicated than in the case of the $T^2$ model but it is numerically analysable. To find an effective potential from (\ref{eqt4}), we still need to substitute $z\to -\frac{g}{16\pi^2}\log{g v_2/\mu^2}$, and in the case of the $T^4$ model
\begin{equation}
    v_2=-\frac{2}{3}\omega ^2 \left(\cosh \left(\sqrt{\frac{2}{3}} \omega  \phi \right)-4\right) \tanh ^2\left(\frac{\omega  \phi }{\sqrt{6}}\right) \text{sech}^4\left(\frac{\omega  \phi }{\sqrt{6}}\right).
\end{equation}

Figure \ref{t4} shows a comparison of the classical potential, the one-loop-corrected potential, with the all loop effective potential for various $\mu^2$. Here we also show spontaneous symmetry breaking for which the contribution from the first loop is responsible, and still the quantum effective potential is characterised by a smoother behaviour. Unlike the previous case the effective potential for the $T^4$-model turns to zero at the minimum, though from a certain value of $\mu^2$ there is also symmetry breaking leading to the appearance of the maximum of the effective potential at $\phi=0$, which has the same meaning of instability as in the case of the $T^2$-model. Also as can be seen from the figure \ref{t4}, the $T^4$ potential is much more stable to quantum corrections and almost does not change in a large range of values of the $\mu^2$ parameter. Thus, for estimations of the all-loop effective potential, one can use the classical $T^4$ potential with a one-loop quantum  correction. 

\begin{figure}[h]
\begin{center}
\includegraphics[scale=0.7]{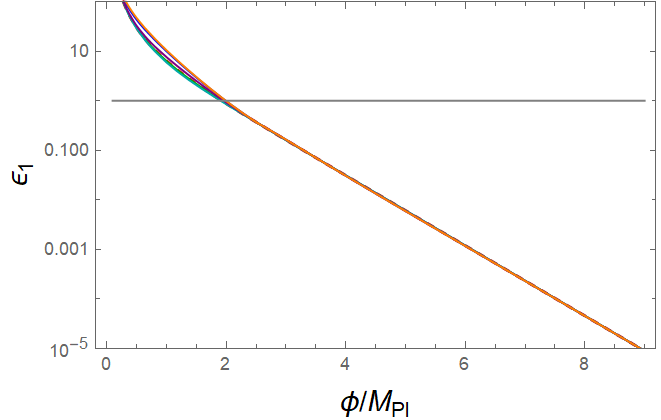}
\caption{Plots for the cosmological parameter $\epsilon_1$.  The red dashed line corresponds to the classical potential, the blue line all-loop effective potential --- $\mu^2 \sim 10^{-2}$, the green $\mu^2 \sim 10^{-10}$, the purple $\mu^2 \sim 10^{-16}$, the orange $\mu^2 \sim 10^{-56}$. The grey line represents the end of inflation} \label{epsdifft4}
\end{center}
\end{figure}

The Fig.  \ref{epsdifft4} shows the cosmological parameter $\epsilon_1$ for several effective potentials. It can be seen that irrespective of the value of the parameter $\mu^2$, the slow rolling is completed (if we do not speak about too small values of the parameter $\mu^2$, where the occurrence of false vacua occurs).

Equations for the effective potentials of $T^{2n}$-models can be derived, but it turns out that they are not fundamentally different from the case of the full RG-summed potential of the $T^4$-model.

\section*{Conclusion}

In this paper, we have constructed quantum effective potentials and used them to study inflationary cosmology with slow roll.  We derived the generalised RG equation for the effective potential in the leading logarithmic approximation and applied it to evaluate the potentials of the $T^2$ and $T^4$-models, which are  often used in modern cosmology of slow roll inflation.  We found  that while the one-loop correction strongly affects the potential breaking its original symmetry, the contribution of higher loops smoothes the behaviour of the potential. However, unlike the $\phi^4$-case ~\cite{CW}, we found that the effective potentials preserve spontaneous symmetry breaking when summing all the leading corrections (\ref{fig:oneloop}). This may lead to effects related to the decay of the metastable vacuum.

As shown in \cite{Kallosh:2016gqp}, quantum one-loop corrections do not deviate the asymptotic values of the effective potential at large fields and can change the form of the potential only at small values of the inflaton field. 
However,  taking into account all leading logarithms by means of the RG-formalism (or generalised RG) may lead to a significant change of the asymptotics~\cite{CW, we2023}. In the case of $T$-models of $\alpha$-attractors, we have found that the asymptotics of the potential remains unchanged when all-loop corrections are taken into account, thus confirming the conclusions of \cite{Kallosh:2016gqp}. 

\begin{figure}[h]
\begin{center}
\includegraphics[scale=0.33]{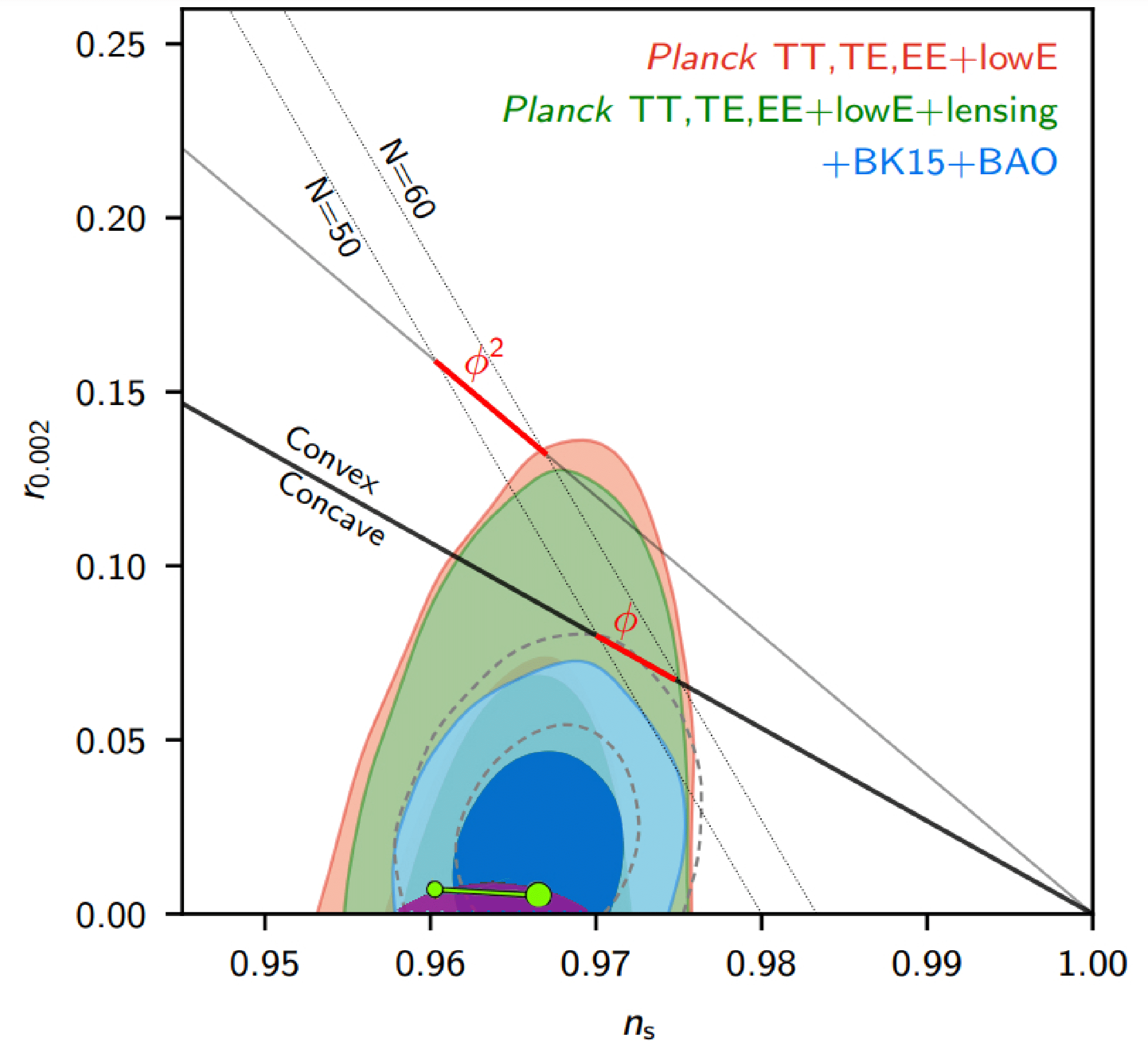}
\caption{Comparison in the  $n_s$ versus $r$ plot. The dark purple region corresponds to the full effective potentials for $T^2$ in the limits $10^2<\mu^2 < 10^{-3}$ and $T^4$ in the limits $10^{10}<\mu^2 < 10^{-14}$ in the interval from 50 to 60 $e$-folds. The bright green line corresponds to the classical $T^2$-model. Data source is taken from \cite{PLANCK} }\label{nsrt2}
\end{center}
\end{figure}

In this work, we also constructed and studied the Hubble flow parameters reflecting the basic properties of slow-roll inflation. It was found that for many values of the dimensional regularization parameter $\mu$ the inflation can become eternal. In some region of parameters the all-loop quantum effective potential only slightly differs from the classical one and leads to a similar slow-roll behavior.
Indeed, our calculations of cosmological inflation parameters fit the diagram of the $n_s-r_s$ plot. In Fig. \ref{nsrt2}, we combined the data for the indices $n_s$ and $r_s$ for the effective potentials of both models in the purple region. The diagram reveals that all effective potentials are consistent with the observational data~\cite{PLANCK}. The purple region on the plot corresponds to all allowed values of the parameter $\mu^2$ for the potential. It can be seen that the effective potentials based on $T^2$ or $T^4$ do not differ significant but are still distinguishable. The $T^2$ and $T^4$ with one-loop corrections can also enter this region, but a fine tuning of the $\mu$ parameter is required (for example, $\mu^2\simeq 2$ for the $T^2$-model in our units).

The approach used in this paper allows one to calculate  the effective potentials for other cosmological models of inflation. It can be used  to constrain or to generalize some inflationary models in the slow-roll approximation.   

\section*{Acknowledgments}
The authors are grateful to A.Starobinsky and S. Ketov for the statement of the problem and for D.Gorbunov for numerous useful discussions of inflationary scenarios. Financial support from the Russia Scientific Foundation Grant \# 21-12-00129 is cordially acknowledged.

\bibliographystyle{unsrt}
\bibliography{refs}

\end{document}